\documentclass[12pt]{article}
\usepackage{amssymb,amsmath,epsfig}
\allowdisplaybreaks
\begin{document}
\title{\bf Noether Symmetries and Some Exact Solutions in $f(R, T^{2})$ Theory}
\author{M. Sharif \thanks {msharif.math@pu.edu.pk} and M. Zeeshan Gul
\thanks{mzeeshangul.math@gmail.com}\\
Department of Mathematics and Statistics,\\ The University of Lahore
54792, Pakistan.}

\date{}
\maketitle

\begin{abstract}
The main objective of this article is to examine some physically
viable solutions through the Noether symmetry technique in $f(R,
T^{2})$ theory. For this purpose, we assume a generalized
anisotropic and homogenous spacetime that yields distinct cosmic
models. In order to investigate Noether equations, symmetry
generators and conserved quantities, we use a specific model of this
modified theory. We find exact solutions and examine the behavior of
various cosmological quantities. It is found the behavior these
quantities is consistent with current observations indicating that
this theory describes the cosmic accelerated expansion. We conclude
that generators of Noether symmetry and conserved quantities exist
in this theory.
\end{abstract}

\section{Introduction}

The current cosmic expansion has been the most stunning and dazzling
result for the scientific community \cite{1}. Although general
relativity (GR) is a widely accepted theory which explains the cause
of this expansion but it has some issues like coincidence and fine
tuning problems. To addresses these issues, several modifications of
GR (modified gravitational theories) have been formulated to unveil
the cosmic mysteries. The first modification of GR is $f(R)$  theory
and significant literature \cite{2} is available to understand the
physical features of this theory. Nojiri and Odinstov \cite{3}
The minimal coupled theory was established
in \cite{4} named as $f(G,T)$ theory while
$f(R,T,R_{\xi\eta}T^{\xi\eta})$ is a non-minimal coupled theory
\cite{5}.

Nojiri and Odintsov \cite{5a} reviewed various modified
gravitational theories such as $f(R)$, $f(G)$ and $f(R,G)$ gravity
models which are assumed as alternative theories for dark energy.
They showed that some of such theories may pass the solar system
tests and have quite rich cosmological structure: they may naturally
describe the effective (cosmological constant, quintessence or
phantom) with a possible transition from deceleration to
acceleration era. The possibility to explain the coincidence problem
as the manifestation of the universe expansion in such models is
mentioned. The late (phantom or quintessence) universe filled with
dark fluid with inhomogeneous equation of state (where inhomogeneous
terms are originated from the modified gravity) is also described.

Nojiri et al \cite{5b} discussed some standard issues and also the
latest developments of modified theories in cosmology, emphasizing
on inflation, bouncing cosmology and late-time acceleration era. In
particular, they presented the formalism of $f(R)$, $f(G)$ and
$f(T)$ gravity theories. They emphasized on the formalism of these
theories and explained how these theories can be considered as
viable descriptions for our universe. They demonstrated how bouncing
cosmology can actually be described by these theories. Moreover,
they discussed several qualitative features of the dark energy era
by using the modified gravity formalism and also discussed how a
unified description of inflation with dark energy era can be
described by using the modified gravity framework. They also
discussed some astrophysical solutions in the context of modified
gravity and several qualitative features of these solutions.
Capozziello and Laurentis \cite{5c} examined that extended theories
of gravity can be considered a new paradigm to cure shortcomings of
GR at infrared and ultraviolet scales.  They focused
on specific classes of theories like $f(R)$ gravity, scalar-tensor
gravity in the metric and Palatini approaches. A number of viability
criteria are presented considering the post-Newtonian and the post
Minkowskian limits. In particular, they discussed the problems of
neutrino oscillations and gravitational waves in extended gravity.

Recently, Katirci and Kavuk \cite{6} modified $f(R)$ theory by
introducing a non-linear term $(T^{2}=T_{\xi\eta}T^{\xi\eta})$ in the
functional action referred to as $f(R,T^{2})$ theory. This
curvature-matter coupled theory develops a particular relation
between geometry and matter which provides a more comprehensive
description to unveil the dark components of the universe. This
proposal is also dubbed as energy-momentum squared gravity (EMSG)
and contains higher-order matter source terms which are helpful to
analyze various interesting cosmological results. It is worthwhile
to mention here that this theory explains the complete cosmic
history and the cosmic evolution.  Roshan and Shojai \cite{7}
examined that EMSG resolves the primordial singularity as it has
bounce in the early universe. Board and Barrow \cite{8} used a
specific model of this theory and discussed exact solution,
singularities as well as cosmic evolution with the isotropic
configuration of matter in this theory. Bahamonde et al \cite{9}
studied various EMSG models and analyzed that these models manifest
the current cosmic evolution and acceleration. Ranjit et al
\cite{10} examined possible solutions for matter density and
discussed their cosmological results in EMSG. We have examined some
physically viable solutions in this theory \cite{11}. We have also
studied the dynamics of celestial objects and found that collapse
rate reduces in EMSG as compared to GR \cite{11b}. Recently, Sharif
and Naz \cite{11c} studied physical characteristics of a gravastar
in this framework.

The universe is homogeneous and isotropic at large scales according
to the cosmic observations such as \emph{Planck satellites} and
\emph{Cosmic microwave background}. But, the cosmos was found to be
anisotropic and spatially homogeneous in the past times. The Cosmic
microwave background temperature is used as anisotropy in the
current universe. Bianchi type (BT) spacetimes are the most
important and elegant models that determine the effect of anisotropy
in the early time, i.e., the less anisotropy stops the rapid cosmic
expansion that yields a highly isotropic universe \cite{12}. Several
researchers have analyzed these models in a different context. The
BT-I model with anisotropic matter has been studied in \cite{13} and
investigated that the effective density and EoS parameter describes
the cosmic expansion. The BT-I model with the dominance of DE  has
been investigated in \cite{14} and they examined that DE is
responsible for cosmic expansion. The exact anisotropic solutions in
curvature-matter coupled theory has been studied in \cite{15}.

The features of a mathematical or physical system that do not change
due to some change are determined by symmetry. Exact solutions of
non-linear differential equations have been formulated in a
significant way using symmetry approaches. Symmetry at geometric
level occurs when a system maintains its shape after going through
particular transformations like rotation, reflection or scaling.
Continuous symmetry appears because of continuous change in the
system such as Noether symmetry (NS) that corresponds to the
Lagrangian. The feasible characteristics of the system can be
discussed by defining the Lagrangian which describes energy content
and gives knowledge about the symmetries exist in the system.
However, the most efficient method for establishing a connection
between generators and the system's conserved values is NS technique
\cite{16}. This approach simplifies the system and offers new
solutions for deciphering the mysterious cosmos.

The NS approach is significant as it recovers symmetry generators as
well as some conservation laws of the system \cite{18}. This method
does not deal only with the dynamical solutions but it also provides
some viable conditions to select cosmic models based on recent
observations \cite{19}. Moreover, this method is an important and
useful technique to examine exact solutions by using conserved
values of the system. Conservation laws are the main ingredients to
analyze the distinct physical phenomena. These are the particular
cases of the Noether theorem, according to which every
differentiable symmetry produces conservation laws. The conservation
laws of linear and angular momentum govern the translational and
rotational symmetry of any object. The Noether charges are important
in the literature as they are used to examine various major cosmic
problems in various considerations \cite{20}-\cite{30}. Exact
solutions of the spherical spacetime through NS in $f(R)$ theory has
been obtained in \cite{33}. The stability of spherical and FRW
universe models in the same theory has been investigated in
\cite{34}. Kucukakca et al \cite{35} found analytic solutions for
various universe models in the scalar-tensor theory. In the
framework of $f(R,T)$ theory, the exact viable solutions has been
examined in \cite{36}. Recently, we have analyzed stable modes of
Einstein universe \cite{38} and the geometry of celestial objects in
EMSG \cite{39}.

This manuscript investigates the NS for anisotropic and homogenous
cosmic models such as BT-I, BT-III and Kantowski-Sachs (KS) in the
background of EMSG. We analyze generators of the NS with conserved
quantities and evaluate cosmological solutions for a specific EMSG
model to investigate the cosmic evolution. The format of this
manuscript is planned as follows. Section \textbf{2} studies the
basic formalism of EMSG. Section \textbf{3} provides a detailed
study of the NS approach and derives exact cosmological solutions
which are then discussed through graphs. The summary of the
consequences is given in section \textbf{4}.

\section{Field Equations}

We derive the field equations of the homogeneous and anisotropic
spacetime in this section. The action of EMSG is expressed as
\cite{4}
\begin{equation}\label{1}
A=\int\left(\frac{f(R ,T^{2})}{2\kappa^2}+L_{m}\right)d^4x\sqrt{-g},
\end{equation}
where $\kappa^{2}=1$ and $\mathcal{L} _{m}$ manifest the coupling
constant and Lagrangian of matter, respectively. The corresponding
equations of motion are obtained as
\begin{equation}\label{3}
R_{\xi\eta}f_{R}+g_{\xi\eta}\Box
f_{R}-\nabla_{\xi}\nabla_{\eta}f_{R}-\frac{1}{2}g_{\xi\eta}f
=T_{\xi\eta}-\Theta_{\xi\eta}f_{T^{2}},
\end{equation}
where $\Box=\nabla_{\xi}\nabla^{\xi},~f_{T^{2}}=\frac{\partial
f}{\partial T^{2}},~f_{R}= \frac{\partial f}{\partial R}$ and
\begin{eqnarray}\nonumber
\Theta_{\xi\eta}=-2L_{m}(T_{\xi\eta}-\frac{1}{2}g_{\xi\eta}T)-4
\frac{\partial^{2}L_{m}}{\partial g^{\xi\eta}\partial
g^{\alpha\beta}}T^{\alpha\beta}-TT_{\xi\eta}+2T_{\xi}^{\alpha}
T_{\eta\alpha}.
\end{eqnarray}
Rearranging Eq.(\ref{3}), we have
\begin{equation}\label{4}
G_{\xi\eta}=\frac{1}{f_{R}}(T_{\xi\eta}^{(D)}+T_{\xi\eta})
=T_{\xi\eta}^{eff},
\end{equation}
where $T_{\xi\eta}=(\rho+p)U_{\xi}U_{\eta}+pg_{\xi\eta}$ and
$T_{\xi\eta}^{(D)}$ defines the modified terms of EMSG, represented
as
\begin{equation}\label{5}
T_{\xi\eta}^{(D)}= \frac{1}{2}g_{\xi \eta}(f-Rf_{R})-g_{\xi\eta}
\Box f_{R}+\nabla_{\xi}\nabla_{\eta}f_{R}-\Theta
_{\xi\eta}f_{T^{2}}.
\end{equation}
We assume a generalized spacetime that corresponds to BT-I, BT-III
and KS spacetimes as
\begin{equation}\label{6}
ds^{2}=-dt^{2}+a^{2}(t)dx^{2}+b^{2}(t)(d\theta^{2}+\psi^{2}(\theta)
d\phi^{2}),
\end{equation}
where $\psi(\theta)=\theta,~ \sinh\theta,~ \sin\theta$ satisfying
the relation
\begin{eqnarray}\nonumber
\frac{1}{\psi}\frac{d^{2}\psi}{d\theta^{2}}=-\epsilon.
\end{eqnarray}
For $\epsilon=0,-1,1$, the BT-I, BT-III and KS cosmic models are
obtained. The resulting equations of motion become
\begin{eqnarray}\nonumber
\rho^{eff}&=&\frac{1}{f_{R}}[\rho-\frac{1}{2}f+(3p^{2}+\rho^{2}+4p\rho)
f_{T^{2}}+\epsilon
b^{-2}f_{R}-(\dot{a}a^{-1}+2\dot{b}b^{-1})\\\label{7}
&\times&(\dot{R}f_{RR}+\dot{T}^{2}f_{RT^{2}})+(\ddot{a}a^{-1}+2\ddot{b}
b^{-1}+2\dot{a}\dot{b}a^{-1}b^{-1}+\dot{b}^{2}b^{-2})f_{R}],\\\nonumber
p^{eff}&=&\frac{1}{f_{R}}[p+\frac{1}{2}f+2\dot{b}b^{-1}(\dot{R}f_{RR}+
\dot{T}^{2}f_{RT^{2}})-\epsilon
b^{-2}f_{R}+\ddot{R}f_{RR}+\ddot{T}^{2}\\\nonumber&\times&f_{RT^{2}}
-(\ddot{a}a^{-1}+2\ddot{b}b^{-1}+2\dot{a}\dot{b}a^{-1}b^{-1}+\dot{b}
^{2}b^{-2})f_{R}+\dot{R}^{2}f_{ RRR}+\dot{T}^{2}\\\label{8}&\times&
f_{RT^{2}T^{2}}+2\dot{R}\dot {T}f_{RRT^{2}}],\\\nonumber
p^{eff}&=&\frac{1}{f_{R}}[p+\frac{1} {2}f+(\dot{a}a^{-1}+\dot{b}
b^{-1})(\dot{R}f_{RR}+\dot {T}^{2}f_{RT^{2}})-\epsilon
b^{-2}f_{R}+\ddot{T}^{2}\\\nonumber&\times& f_{RT^{2}}+\ddot{R}f_{R
R}-(\ddot{a}a^{-1}+2\ddot{b}b^{-1}+2\dot{a}\dot{b}a^{-1}b^{-1}+\dot{
b}^{2}b^{-2})f_{R}+\dot{R}^{2}\\\label{9}&\times&f_{RRR}+\dot{T}^{2}
f_{RT^{2}T^{2}}+2\dot{R}\dot{T}f_{RRT^{2}}],
\end{eqnarray}
Now, we apply Lagrange multiplier method to formulate the Lagrangian
as
\begin{eqnarray}\nonumber
L&=&ab^{2}(f-Rf_{R}-T^{2}f_{T^{2}}+(3p^{2}+\rho^{2})f_{T^{2}}+p)-2a
(2\dot{a}\dot{b}ba^{-1}\\\label{11}&+&\dot{b}^{2}-\epsilon)f_{R}
-(2b^{2}\dot{a}+4ab\dot{b}) \dot{R}f_{RR}-(2b^{2}\dot{a}+4ab\dot{b})
\dot{T}^{2}f_{RT^{2}}.
\end{eqnarray}
The fundamental properties of the system can be explained using the
Hamiltonian $(E)$ and the dynamical equations , determined as
\begin{equation}\label{12}
\frac{\partial L}{\partial q^{i}}-\frac{d} {dt}(\frac{\partial
L}{\partial\dot{ q}^{i}})=0, \quad E=\dot{q}^{i}(\frac{\partial
L}{\partial \dot{q}^{i}})-L,
\end{equation}
where generalized coordinates are denoted by $q^{i}$. The resulting
dynamical equations are
\begin{eqnarray}\nonumber
&&f-Rf_{R}-T^{2}f_{T^{2}}+(3 p^{2}+\rho^{2})f_{T^{2}}+p+4\dot{
b}b^{-1}(\dot{R}f_{RR}+f_{RT^{2}}\\\nonumber&&\times \dot{T}^{2})
+a(f_{T^{2}}(6pp_{,a}+2\rho\rho_{,a})+p_{,a})+b^{-2}(2\dot{b}^{2}
+4\ddot{b}b+2\epsilon)f_{R}\\\label{13}&&+2\ddot{R}f_{RR}+2\ddot{T}
^{2}f_{RT^{2}}+2\dot{R}^{2}f_{RRR}+2\dot{T}^{2}f_{RT^{2}T^{2}}+4
\dot{R}\dot{T}^{2}f_{RRT^{2}}=0,\\\nonumber&&f-Rf_{R}-T^{2}f_{T^{2}}
+(3p^{2}+\rho^{2})f_{T^{2}}+(\dot{R}f_{RR}+\dot{T}^{2}f_RT^{2})4\dot
{a}a^{-1}\\\nonumber&&+p+b(f_{T^{2}}(6pp_{,b}+2\rho\rho_{,b})+p_{,b})
+2a^{-1}b^{-1}(\ddot{a}b+\dot{a}\dot{b}+\ddot{b}a)f_{R}\\\label{14}&&
+2\ddot{f_{R}}+4\dot{b}b^{-1}(\dot{R}f_{RRT^{2}}+\dot{T}^{2}f_{RT^{2}T
^{2}})=0,\\\nonumber&&(2\ddot{a}a^{-1}-4\ddot{b }b^{-1})f_{RR}+(3p
^2+\rho^{2})f_{RT^{2}}-2\epsilon b^{-1}f_{RR}-R f_{RR}\\\label{15}&&
-T^{2}f_{RT^{2}}-(4\dot{a}\dot{b}a^{-1}b^{-1}+2\dot{b}^{2}b^{-2})f_{RR}
=0,\\\nonumber&& (2\ddot{a}a^{-1}-4\ddot{b}b^{-1})f_{RT^{2}}+(3p^2+
\rho^{2})f_{T^{2}T^{2}}-2\epsilon b^{-1}f_{RT^{2}}-Rf_{RT^{2}}
\\\label{16}&&-T^{2}f_{T^{2}T^{2}}-(4\dot {a}\dot{b}a^{-1}b^{-1}+2\dot
{b}^{2}b^{-2})f_{ RT^{2}}=0.
\end{eqnarray}
We formulate the Hamiltonian to examine the total energy of the
system as
\begin{eqnarray}\nonumber
E&=&-ab^{2}(f-Rf_{R}-T^{2}f_{T^{2}}+(3p^{2}+\rho^{2})f_{T^{2}})-(2\dot{a}
\dot{b}ba^{-1}+\dot{ b}^{2})\\\label{17}&\times& 2af_{ R}-\epsilon
f_{R}-ab^{2}p -(2b^{2}\dot{a}+4a b\dot{b})\dot{f_{R}}.
\end{eqnarray}
The dynamical equations (\ref{13})-(\ref{16}) are extremely complex
due to metavariable functions and their derivatives. Despite being
very challenging to solve these equations directly, these can be
solved in one of two ways. The first is numerically or analytically
approach, and the second is NS technique to identify exact
solutions. Although this theory is not conserved but one can obtain
conserved values through NS approach, which are then used to examine
the mysterious universe. As a result,  latter strategy because it is
more intriguing. Thus, we will employ the later approach as it is
more interesting.

\section{Noether Symmetries in EMSG}

The NS strategy gives a fascinating method to develop new cosmic
models and associated structures in modified theories of gravity.
This section formulates the Noether equations for the homogenous and
anisotropic universe model in EMSG. This approach establishes a
vector field that corresponds to tangent space. Thus, the vector
field acts as a generator and gives conserved quantities that can be
used to analyze viable solutions. The symmetry generators are
expressed as
\begin{equation}\nonumber
Y=\lambda(t,q^{i})\frac{\partial}{\partial t}+\Upsilon^{j}(t,
q^{i})\frac{\partial}{\partial q^{j}},\quad i=1,2,3...,n,
\end{equation}
where $\lambda(t,a, b, R,T^{2})$ and $\Upsilon^{j}(t,a, b, R,T^{2})$
are the unknown parameters. In our case, the Lagrangian admits five
degrees of freedom and hence the symmetry generator becomes
\begin{equation}\label{18}
Y= \lambda\frac{\partial}{\partial
t}+\Upsilon^{1}\frac{\partial}{\partial a}
+\Upsilon^{2}\frac{\partial}{\partial b}
+\Upsilon^{3}\frac{\partial}{\partial R}
+\Upsilon^{4}\frac{\partial}{\partial T^{2}}.
\end{equation}
The Lagrangian must satisfy the invariance constraint, expressed as
\begin{eqnarray}\label{19}
Y^{[1]}L+(D\lambda)L= D\Omega, \quad
Y^{[1]}=Y+\dot{\Upsilon}^{i}\frac{\partial} {\partial\dot{q}^{i}}
\end{eqnarray}
where $\Omega$ is the boundary term and $D=\frac{\partial}{\partial
t}+\dot{ q}^{i}\frac{\partial}{\partial q^{i}}$ defines the total
derivative. The corresponding integral integral of motion is
expressed as
\begin{equation}\label{20}
I=\Upsilon^{i}\frac{\partial L}{\partial\dot{ q}^{i}}-\lambda
E-\Omega.
\end{equation}
This is a crucial component of NS that is essential for computing
viable solutions and is also named as the conserved quantities.

We take the vector field $(Y)$ with configuration space $Q= ({t,a,
b, R,T^{2}})$ to examine the generators with corresponding first
integrals of Lagrangian (\ref{11}) under invariance condition
(\ref{19}). By comparing the coefficients of Eq.(\ref{19}), we have
\begin{eqnarray}\label{21}
&&2b^{2}\Upsilon^{1}_{,t}f_{RT ^{2}}+4ab\Upsilon^{2}_{,t}f_{R
T^{2}}+\Omega_{,T^{2}}=0,
\\\label{22}&&
2b^{2}\Upsilon^{1}_{,t}f_{RR} +4ab\Upsilon^{2}_{,t}f_{R
R}+\Omega_{,R}=0, \quad \lambda_{,a}f_{R}=0,
\\\label{23}&&
b\Upsilon^{1}_{,T^{2}}f_{RR} +b\Upsilon^{1}_{,R}f_{RT^{2}}
+2a\Upsilon^{2}_{,T^{2}}f_{RR} +2a\Upsilon^{2}_{,R}f_{RT^{2}} =0,
\\\label{24}&&
4b\Upsilon^{1}_{,t}f_{R}+4a \Upsilon^{2}_{,t}f_{R}+4ab
\Upsilon^{3}_{,t}f_{RR}+4a b\Upsilon^{4}_{,t}f_{RT^{2}}
+\Omega_{,b}=0,
\\\label{25}&&
4b\Upsilon^{2}_{,t}f_{R}+2b ^{2}\Upsilon^{3}_{,t}f_{RR}+2b
^{2}\Upsilon^{4}_{,t}f_{RT^{2}}+\Omega_{ ,a}=0, \quad
\lambda_{,b}f_{R}=0,
\\\label{26}&&
b\Upsilon^{1}_{,T^{2}}f_{RR} +2ab\Upsilon^{2}_{,T^{2}}f_{
RT^{2}}=0,\quad \lambda_{,R}f_{RR}=0, \quad
\lambda_{,T^{2}}f_{RT^{2}}=0,
\\\label{27}&&
b\Upsilon^{1}_{,R}f_{RR} +2ab\Upsilon^{2}_{,R}f_{R R}=0, \quad
2\Upsilon^{2}_{,a}f_{R}+b\Upsilon^{3} _{,a}f_{RR}+b\Upsilon^{4}
_{,a}f_{RT^{2}}=0,
\\\nonumber&&
\Upsilon^{1}f_{R}+a\Upsilon^{3}f_{R R}+a\Upsilon^{4}f_{RT^{2}}
+2b\Upsilon^{1}_{,b}f_{R}+2a b\Upsilon^{4}_{,b}f_{RT^{2}}
+2ab\Upsilon^{3}_{, b}f_{RR}
\\\label{28}&&
-a\lambda_{,t}f_{R}+2a \Upsilon^{2}_{,b}f_{R}=0, \quad
\lambda_{,a}f_{RR}=0, \quad \lambda_{,a}f_{RT^{2}}=0,
\\\nonumber&&
+2\Upsilon^{2}_{,R}f_{R}+b\Upsilon^{3} _{,R}f_{RR}+b\Upsilon^{4}
_{,R}f_{RT^{2}}-b\lambda _{,t}f_{RR}+2\Upsilon^{2}f_{R
R}+b\Upsilon^{1}_{,a}f_{R R}
\\\label{29}&&
+b\Upsilon^{3}f_{RRR} +b\Upsilon^{4}f_{RRT^{2}}
+2a\Upsilon^{2}_{,a}f_{RR} =0, \quad \lambda_{,b}f_{RR}=0,
\\\nonumber&&
2\Upsilon^{2}f_{RT^{2}}-b\lambda _{,t}f_{RT^{2}}+b
\Upsilon^{3}_{,T^{2}}f_{RR} +b\Upsilon^{1}_{,a}f_{R
T^{2}}+b\Upsilon^{4}_{,T^{2}} f_{RT^{2}}+2\Upsilon^{2}_{,T^{2}}
f_{R}
\\\label{30}&&
+b\Upsilon^{4}f_{RT^{2}T^{2}} +2a\Upsilon^{2}_{,a}f_{RT^{2}}
+b\Upsilon^{3}f_{RRT^{2}}=0, \quad \lambda_{,b}f_{RT^{2}}=0,
\\\nonumber&&
2\Upsilon^{2}f_{R}+2b\Upsilon^{3}f_{R R}+2b\Upsilon^{4}f_{RT^{2}}
+2b\Upsilon^{1}_{,a}f_{R}+2a \Upsilon^{2}_{,a}f_{R}+2ab
\Upsilon^{3}_{,a}f_{RR}
\\\label{31}&&
+2b\Upsilon^{2}_{,T^{2}}f_{R}+b ^{2}\Upsilon^{3}_{,b}f_{RR}+2a
b\Upsilon^{4}_{,a}f_{RT^{2}} +b^{2}\Upsilon^{4}_{,b}f_{RT^{2}}
-2b\lambda_{,t}f_{R}=0,
\\\nonumber&&
2b\Upsilon^{1}f_{RR}+2a\Upsilon^{2} f_{RR}+2ab\Upsilon^{3}f_{R
RR}+2ab\Upsilon^{4}f_{R RT^{2}}+b^{2}\Upsilon^{1}_{,b} f_{RR}
\\\nonumber&&
+2b\Upsilon^{1}_{,R}f_{R}+2a b\Upsilon^{2}_{,b}f_{RR} +2a
\Upsilon^{2}_{,R}f_{R}+2a b\Upsilon^{3}_{,R}f_{RR}
+2ab\Upsilon^{4}_{,R}f_{R T^{2}}
\\\label{32}&&
-2ab\lambda_{,t}f_{RR}=0, \quad \lambda_{,R}f_{R}=0, \quad
\lambda_{,T^{2}}f_{R}=0,
\\\nonumber&&
2b\Upsilon^{1}f_{RT^{2}}+2a\Upsilon^{2}
f_{RT^{2}}+2ab\Upsilon^{3}f_{R RT^{2}}+2ab\Upsilon^{4}f_{R
T^{2}T^{2}}+b^{2}\Upsilon^{1}_{,b} f_{RT^{2}}
\\\nonumber&&
+2b\Upsilon^{1}_{,T^{2}}f_{R}+2a b\Upsilon^{2}_{,b}f_{RT^{2}}
+2a\Upsilon^{2}_{,T^{2}}f_{R}+2a b\Upsilon^{3}_{,T^{2}}f_{RR}
+2ab\Upsilon^{4}_{,T^{2}}f_{R T^{2}}
\\\label{33}&&
-2ab\lambda_{,t}f_{R T^{2}}=0, \quad \lambda_{,R}f_{RT^{2}}=0, \quad
\lambda_{,T^{2}}f_{RR}=0,
\\\nonumber&&
b^{2}\Upsilon^{1}[f-Rf_{R} -T^{2}f_{T^{2}}+(3p^{2}+\rho^{2})
f_{T^{2}}+p+a((6p p_{,a}+2\rho\rho_{,a}) f_{T^{2}}
\\\nonumber&&
+p_{,a})+2\epsilon f_{R}]+\Upsilon^{2}[2ab(f- Rf_{R}-T^{2}f_{T^{2}}
+(3p^{2}+\rho^{2})f_{T^{2}}+p)
\\\nonumber&&
+ab^{2}((6pp_{, b}+2\rho\rho_{,b})f_{T ^{2}}+p_{,b})]+(3p^{2}+\rho^{
2})f_{RT^{2}})+\Upsilon^{3}[-a b^{2}(Rf_{RR}
\\\nonumber&&
-T^{2}f_{RT^{2}}+(3p ^{2}+\rho^{2})f_{T^{2}T^{2}})+2 a\epsilon
f_{RR}]+\Upsilon^{4} [-ab^{2}(Rf_{R T^{2}}-T^{2}f_{T^{2}T^{2}}
\\\nonumber&&
-(3p^{2}+\rho^{2})f_{T^{2}T^{2}}) +2a\epsilon f_{RT^{2}}]
+\lambda_{,t}[ab^{2}(f-Rf_{R}-T^{2}f_{T^{2}}+2\epsilon f_{R}
\\\label{34}&&
+(3p^{2}+\rho^{2})f_{T^{2}}+p)] -\Omega_{,t}=0.
\end{eqnarray}
These equations help to study the dark cosmos in the context of
$f(R,T^{2})$. We solve the above system to obtain exact solutions
for specific $f(R ,T^{2})$ model in the following section.

\subsection{Exact Solutions}

Here, we formulate the generators of NS, conserved values of the
system and corresponding physical solutions. Due to the  above
system's complexity and nonlinearity, we assume a particular EMSG
model $f(R,T^{2})= R+T^{2}$ which minimizes the system complexity
and help to examine the exact solutions \cite{5}. Manipulating
Eqs.(\ref{21})-(\ref{34}), we obtain
\begin{eqnarray}\nonumber
\Upsilon^{1}&=&\frac{1}{3}a\dot{F_{1}}(t)-2c_{1}a-\frac{1}{2}\frac{aF_{2}
(t)}{b^{\frac{3}{2}}}-\frac{3}{8}\frac{F_{4} (t)}{\sqrt
b}+\frac{c_{2}}{\sqrt{b}},\\\nonumber\Upsilon^{2}&=&\frac{F_{2}(t)}
{\sqrt{b}}+(\frac{1}{3}\dot{F_{1}}(t)+c_{1}) b,\quad
\lambda=F_{1}(t),\\\nonumber\Psi&=&-\frac{4}{3}ab^{2}\ddot{F_{1}}(t)
-4a\sqrt{b}\dot{F_{2}}(t)+F_{3} (t)+\dot{F_4}(t)b^{\frac{3}{2}},
\\\label{36}\rho&=&\frac{\sqrt{3c_{1}(-3c_{1}a^{2}\epsilon-6c_{1}a
\epsilon -2c_{3}at)}}{3c_{1} ba},
\end{eqnarray}
where $c_{i} (i=1,...,5)$ are integration constants with
$c_{1}\neq0$. The corresponding symmetry generators become
\begin{eqnarray}\nonumber
Y_{1}=-3t\frac{\partial}{\partial t}, \quad
Y_{2}=-3a\frac{\partial}{\partial a}.
\end{eqnarray}
Substituting the value of Lagrangian (\ref{11}), Hamiltonian
(\ref{17}) and above solutions (\ref{36}) in Eq.(\ref{20}), we
obtain first integral as
\begin{eqnarray}\nonumber
I=12ab\dot{b}c_{1}+3 \bigg[\frac{3c_{1}a^{2}\epsilon+6c_{1}
a\epsilon+2c_{3}at}{3c_{1}a}-2a\dot{b}^{2}-4\dot{a}\dot{b}b
+2\frac{\epsilon}{b^{2}}\bigg]c_{1}t-c_{3}t^{2}.
\end{eqnarray}
By comparing the coefficients of $c_{1}$ and $c_{3}$, we have
\begin{eqnarray}\nonumber
I_{1}=t^{2}, \quad I_{2}&=&12ab\dot{b}+3t (\epsilon
a-2a\dot{b}^{2}-4\dot{a} \dot{b}b+\frac{2\epsilon}{b^{2}}).
\end{eqnarray}
We substitute Eqs.(\ref{36}) into dynamical equations
(\ref{13})-(\ref{17}) and obtain the exact solution as
\begin{eqnarray}\nonumber
a(t)&=&\frac{6c_{5}c_{3} (c_{5}+t)^{\frac{2}{3}}-15c_{1}\epsilon
(c_{5}+t)^{\frac{2}{3}}-4c_{3}t (c_{5}+t)^{\frac{2}{3}}+60c_{4}
c_{1}\epsilon}{60c_{1}\epsilon(c_{5}+t)^{\frac{2}{3}}},
\\\label{39}b(t)&=&\frac{1}{2}\sqrt{-6\epsilon}(c_{5}+t).
\end{eqnarray}

To analyze this solution, we study the graphical behavior of some
important cosmological parameters like \emph{deceleration} and $r-s$
parameters that are the major factors in the field of cosmology. The
Hubble parameter $(H)$ defines the cosmic expansion rate that how
quickly it is expanding. The deceleration parameter $(q)$ evaluates
the behavior of cosmic expansion. These cosmic parameters for
anisotropic and homogeneous universe model are defined as
\begin{eqnarray}\nonumber
H=\frac{1}{3}(\frac{\dot{a}}{a}+2\frac{\dot{b}}{b}), \quad
q=-\frac{H}{H^{2}}-1.
\end{eqnarray}
The values of these cosmological parameters become
\begin{eqnarray}\nonumber
H&=&\frac{2}{3}\bigg[4c_{5}c_{3} (c_{5}+t)^{\frac{2}{3}}-15(c_{5}
+t)^{\frac{2}{3}}c_{1}\epsilon-6c _{3}t(c_{5}+t)^{\frac{2}{3}}
+40c_{4}c_{1}\epsilon\bigg]\\\nonumber&\times&\bigg[(6c_{5}c_{3}
(c_{5}+t)^{\frac{2}{3}}-15(c_{5}+t)^{\frac{2}{3}}c_{1}\epsilon
-4c_{3}t(c_{5}+t)^{\frac{2}{3}}+60c_{4}c_{1}\epsilon)(c_{5}+t)\bigg]
^{-1},\\\nonumber q&=&\frac{25}{2}\bigg[160c_{4}^{2}c_{1}^{2}
\sqrt[3]{c_{5}+t}\epsilon^{2}+4c_{5}^{3}c_{3}^{2}(c_{5}+t)^{\frac{2}
{3}}-12c_{5}^{2}c_{1}c_{3}(c_{5}+t)^{\frac{2}{3}}\epsilon+4c_{5}^{2}
c_{3}^{2}\\\nonumber&\times&(c_{5}+t)^{\frac{2}{3}}t+9c_{1}^{2}(c_{5}
+t)^{\frac{2}{3}}\epsilon^{2}c_{5}-12c_{5}c_{1}c_{3}(c_{5}+t)^{\frac{2}
{3}}\epsilon t+9c_{1}^{2}(c_{5}+t)^{\frac{2}{3}}\epsilon^{2}t
\\\nonumber&+&56c_{4}c_{5}^{2}c_{1}c_{3}\epsilon-60c_{4}c_{5}c_{1}^{2}
\epsilon^{2}+72c_{4}c_{5}c_{1}c_{3} \epsilon
t-60c_{4}c_{1}^{2}\epsilon^{2} t+16c_{4}c_{1}c_{3}\epsilon
t^{2}\bigg]\\\nonumber&\times&\bigg[(4c_{5}c_{3}(c_{5}+t)^{\frac{2}{3}}
-15(c_{5}+t)^{\frac{2}{3}}c_{1}\epsilon-6c_{3}t(c_{5}+t)^{\frac{2}{3}}
+40c_{4}c_{1}\epsilon)^{2}\\\nonumber&\times&\sqrt[3]{c_{5}+t}\bigg]^{-1}.
\end{eqnarray}
The pair of $r-s$ parameters constructs a relationship between
formulated and standard models of the universe which is used to
examine the characteristics of DE, expressed as
\begin{eqnarray}\nonumber
r=q+2q^{2}-\frac{\dot{q}}{H}, \quad s=\frac{r-1}{3(q-\frac{1}{2})}.
\end{eqnarray}
For $(r,s)=(1,0)$, the constructed model corresponds to $\Lambda$CDM
model whereas quintessence and phantom DE eras are obtained for
$s>0$ and $r<1$, respectively. The values of these parameters are
bestowed in appendix \textbf{A}. The effective matter variables turn
out to be
\begin{eqnarray}\nonumber
\rho^{eff}&=&\bigg[(6c_{5}c_{3}(c_{5}+t) ^{\frac{2}{3}}-15\epsilon
c_{1}(c_{5}+t)^{\frac{2}{3}} -4c_{3}t(c_{5}+t)^{\frac{2}{3}}+60
c_{4}\epsilon
c_{1})(c_{5}+t)^{2}\bigg]^{-1}\\\nonumber&\times&\bigg[\bigg\{
\frac{1}{\epsilon(c_{5}+t)^{\frac{4}{3}}}(6c_{5}c_{3}(c_{5}+t)
^{\frac{2}{3}}-15c_{1}\epsilon(c_{5}+t)^{\frac{2}{3}}-4c_{3}t
(c_{5}+t)^{\frac{2}{3}}+60\\\nonumber&\times&c_{4}\epsilon
c_{1})(2c_{5}c _{3}(c_{5}+t)^{\frac{2}{3}}+35c_{1}\epsilon(c
_{5}+t)^{\frac{2}{3}}+12c_{3}t(c_{5}+t)^{\frac{2}{3}}+20c_{4}
\epsilon
c_{1})\bigg\}^{\frac{1}{2}}\\\nonumber&\times&\sqrt{-2\epsilon}
(c_{5}+t)^{\frac{2}{3}}c_{5}+\bigg[\frac{1}{\epsilon(c_{5}+t)
^{\frac{4}{3}}}(6c_{5}c_{3}(c_{5}+t)^{\frac{2}{3}} -15\epsilon
c_{1}(c_{5}+t)^{\frac{2}{3}}\\\nonumber&-&4c_{3}t(c_{5}+t)^{
\frac{2}{3}}+60c_{4}\epsilon
c_{1})(2c_{5}c_{3}(c_{5}+t)^{\frac{2}{3}}+35\epsilon
c_{1}(c_{5}+t)^{\frac{2}{3}} +20c_{4}\epsilon
c_{1}\\\nonumber&+&12c_{3}t(c_{5}+t)^{\frac{2}{3}})\bigg]^{\frac
{1}{2}}\sqrt{-2\epsilon}(c_{5}+t)^{\frac{2}{3}}t+2c_{5}c_{3}
(c_{5}+t)^{\frac{2}{3}}+35\epsilon
c_{1}(c_{5}+t)^{\frac{2}{3}}\\\nonumber&+&12c_{3}t(c_{5}
+t)^{\frac{2}{3}}+20c_{4}\epsilon c_{1}\bigg],\\\nonumber
p^{eff}&=&\bigg[2c_{5}c_{3}(c_{5}+t) ^{\frac{2}{3}}+35\epsilon
c_{1}(c_{5}+t)^{\frac{2}{3}} +12c_{3}t(c_{5}+t)^{\frac{2}{3}}+20c
_{4}\epsilon c_{1}\bigg]\bigg[(6c_{5}c_{3}\\\nonumber&\times&
(c_{5}+t)^{\frac{2}{3}}-15\epsilon
c_{1}(c_{5}+t)^{\frac{2}{3}}-4c_{3}t(c_{5}+t)^{\frac{2}{3}}+60
c_{4}\epsilon c_{1})(c_{5}+t) ^{2}\bigg]^{-1}.
\end{eqnarray}
The EoS parameter $(\omega^{eff}= \frac{p^{eff}}{\rho^{eff}})$ is a
dimensionless quantity that determines the correlation between state
parameters. This parameter differentiates the DE era into
quintessence $(-1<\omega\leq -1/3)$ and phantom $(\omega<-1)$
phases. This is given as
\begin{eqnarray}\nonumber
\omega^{eff}&=&\bigg[(2c_{5}c_{3}(c_{5}+t)^{\frac{2}{3}}+35c_{1}
\epsilon(c_{5}+t)^{\frac{2}{3}}+12c_{3}t(c_{5}+t)^{\frac{2}{3}}
+20c_{4}c_{1}\epsilon)\\\nonumber&\times&\bigg\{\bigg[\frac{1}{\epsilon
(c_{5}+t)^{\frac{4}{3}}}\bigg\{(6c_{5}c_{3}(c_{5}+t)^{\frac{2}{3}}
-15c_{1}\epsilon(c_{5}+t)^{\frac{2}{3}}-4c_{3}t(c_{5}+t)^{\frac{2}{3}}
\\\nonumber&+&60c_{4}c_{1}\epsilon)(2c_{5}c_{3}(c_{5}+t)^{\frac{2}{3}}
+35c_{1}\epsilon(c_{5}+t)^{\frac{2}{3}}+12c_{3}t(c_{5}+t)^{\frac{2}{3}}
\\\nonumber&+&20c_{4}c_{1}\epsilon)\bigg\}\bigg]^{\frac{1}{2}}\sqrt
{2}\sqrt{-\epsilon}(c_{5}+t)^{\frac{2}{3}}c_{5}-\bigg[\frac{1}{\epsilon
(c_{5}+t)^{\frac{4}{3}}}\bigg\{(6c_{5}c_{3}(c_{5}+t)^{\frac{2}{3}}
\\\nonumber&-&15c_{1}\epsilon(c_{5}+t)^{\frac{2}{3}}-4c_{3}t(c_{5}+t)
^{\frac{2}{3}}+60c_{4}c_{1}\epsilon)(2c_{5}c_{3}(c_{5}+t)^{\frac{2}{3}}
+35c_{1}\epsilon\\\nonumber&\times&(c_{5}+t)^{\frac{2}{3}}+12c_{3}t(c_{5}
+t)^{\frac{2}{3}}+20c_{4}c_{1}\epsilon)\bigg\}\bigg]^{\frac{1}{2}}\sqrt{2}
\sqrt{-\epsilon}(c_{5}+t)^{\frac{2}{3}}t\\\nonumber&-&2c_{5}c_{3}
(c_{5}+t)^{\frac{2}{3}}-35c_{1}\epsilon(c_{5}+t)^{\frac{2}{3}}-12c_{3}t
(c_{5}+t)^{\frac{2}{3}}-20c_{4}c_{1}\epsilon\bigg\}^{-1}\bigg].
\end{eqnarray}
\begin{figure}\center
\epsfig{file=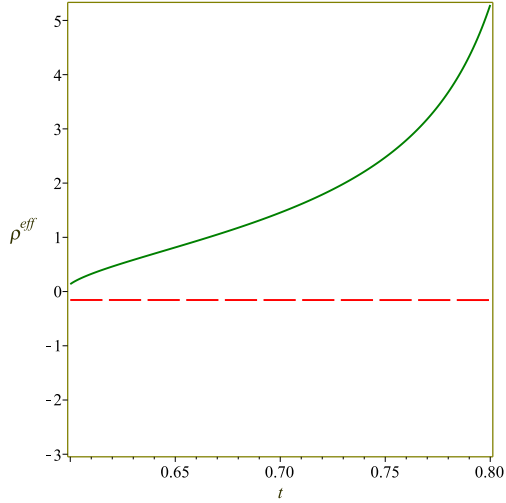,width=.5\linewidth} \caption{Behavior of
effective energy density for $\epsilon=-1$ (line 1) and $\epsilon=1$
(line 2).}
\end{figure}
\begin{figure}
\epsfig{file=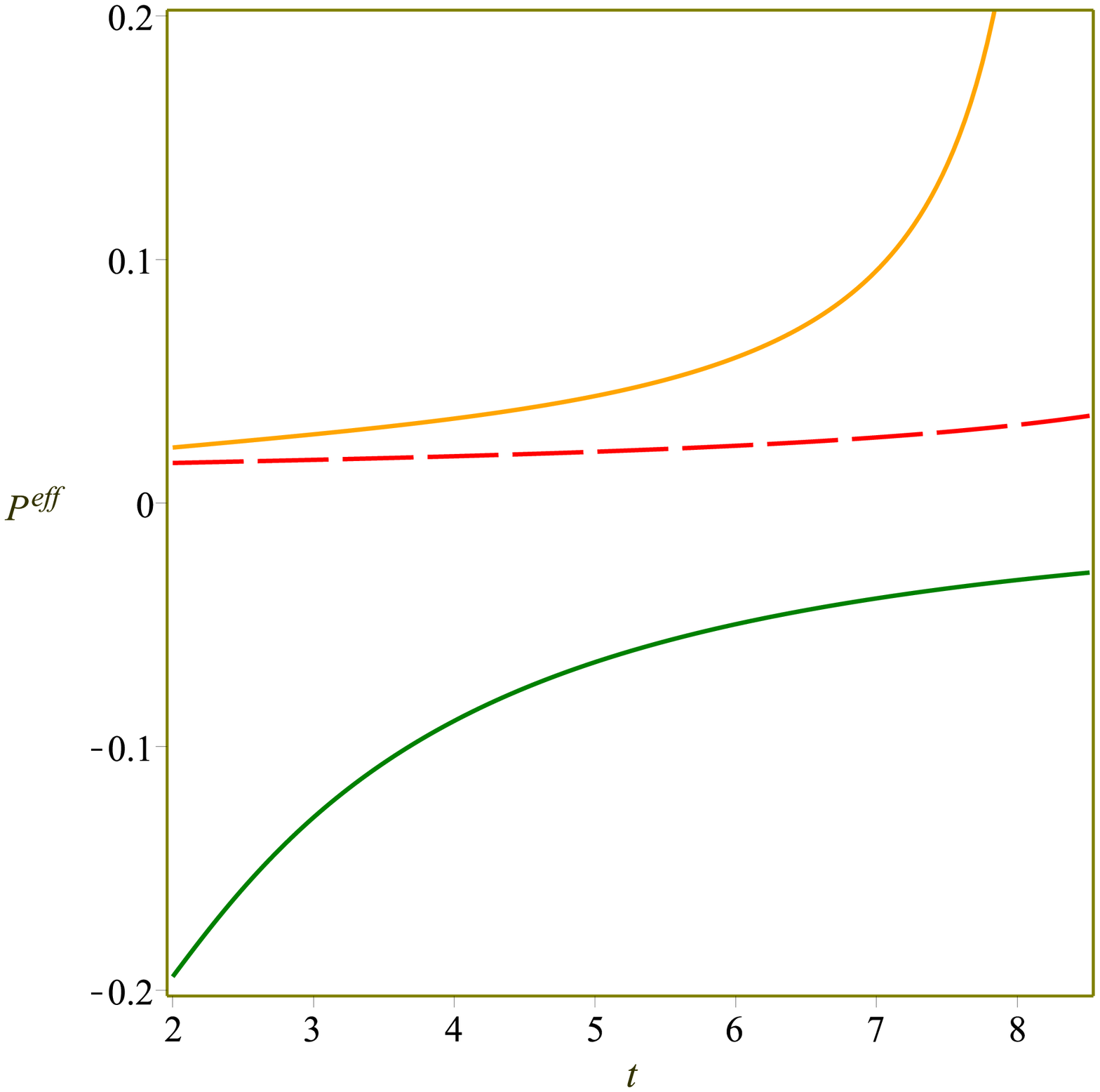,width=.5\linewidth}
\epsfig{file=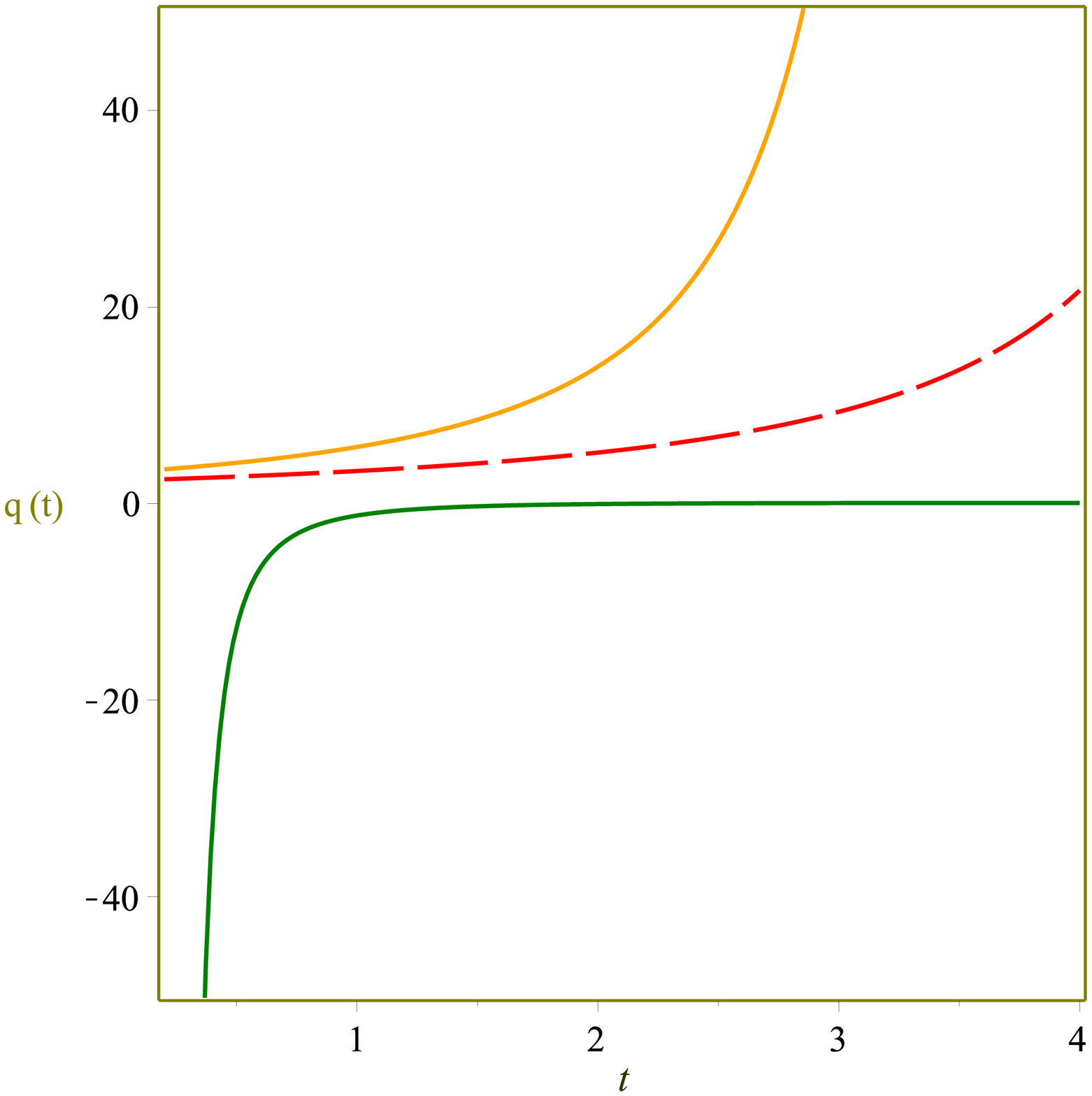,width=.5\linewidth} \caption{Behavior of
effective pressure (Left) and deceleration parameter (Right) for
$\epsilon=-1$ (line 1), $\epsilon=1$ (line 2) and $\epsilon=0$ (line
3).}
\end{figure}
\begin{figure}
\epsfig{file=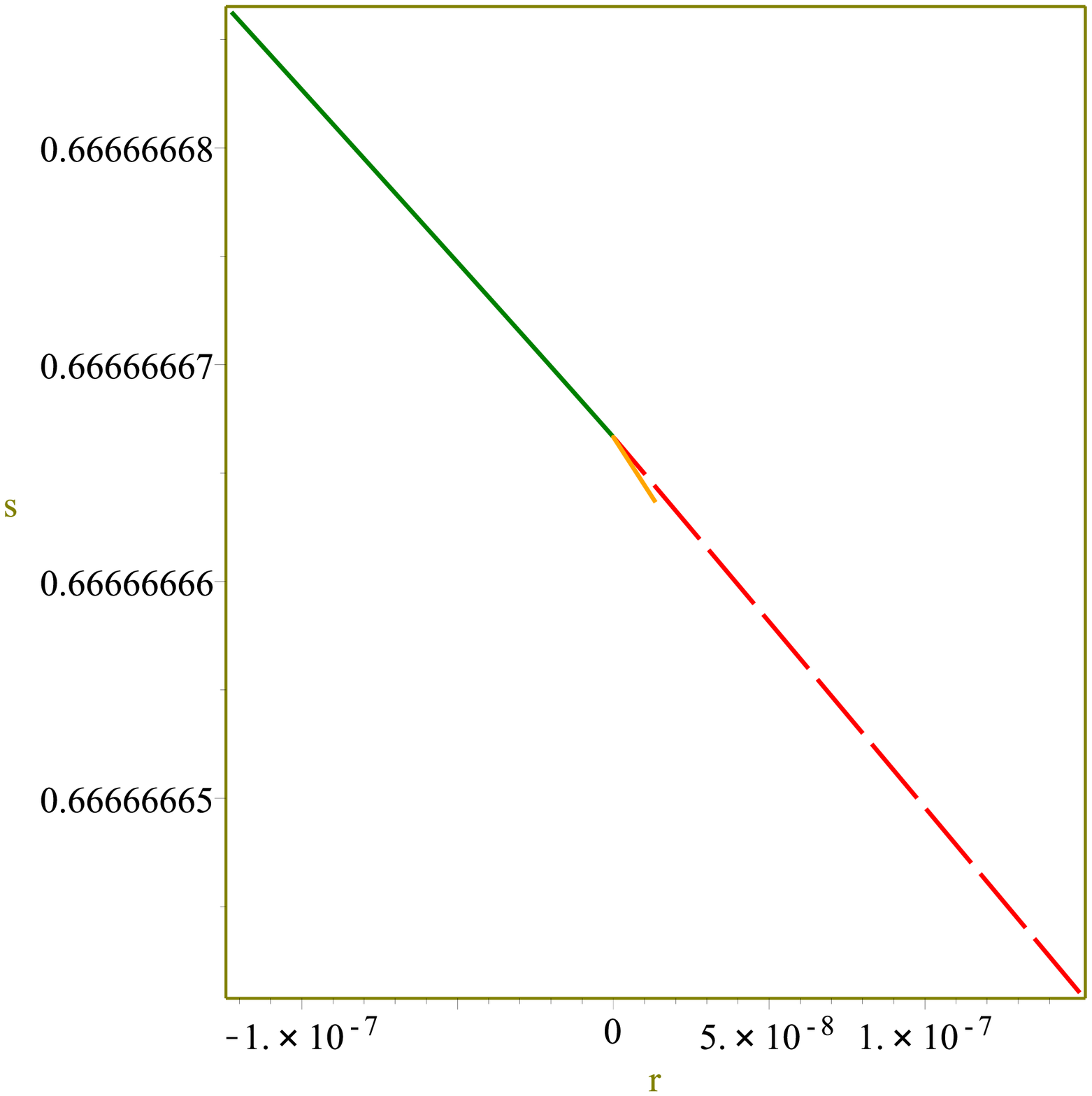,width=.5\linewidth}
\epsfig{file=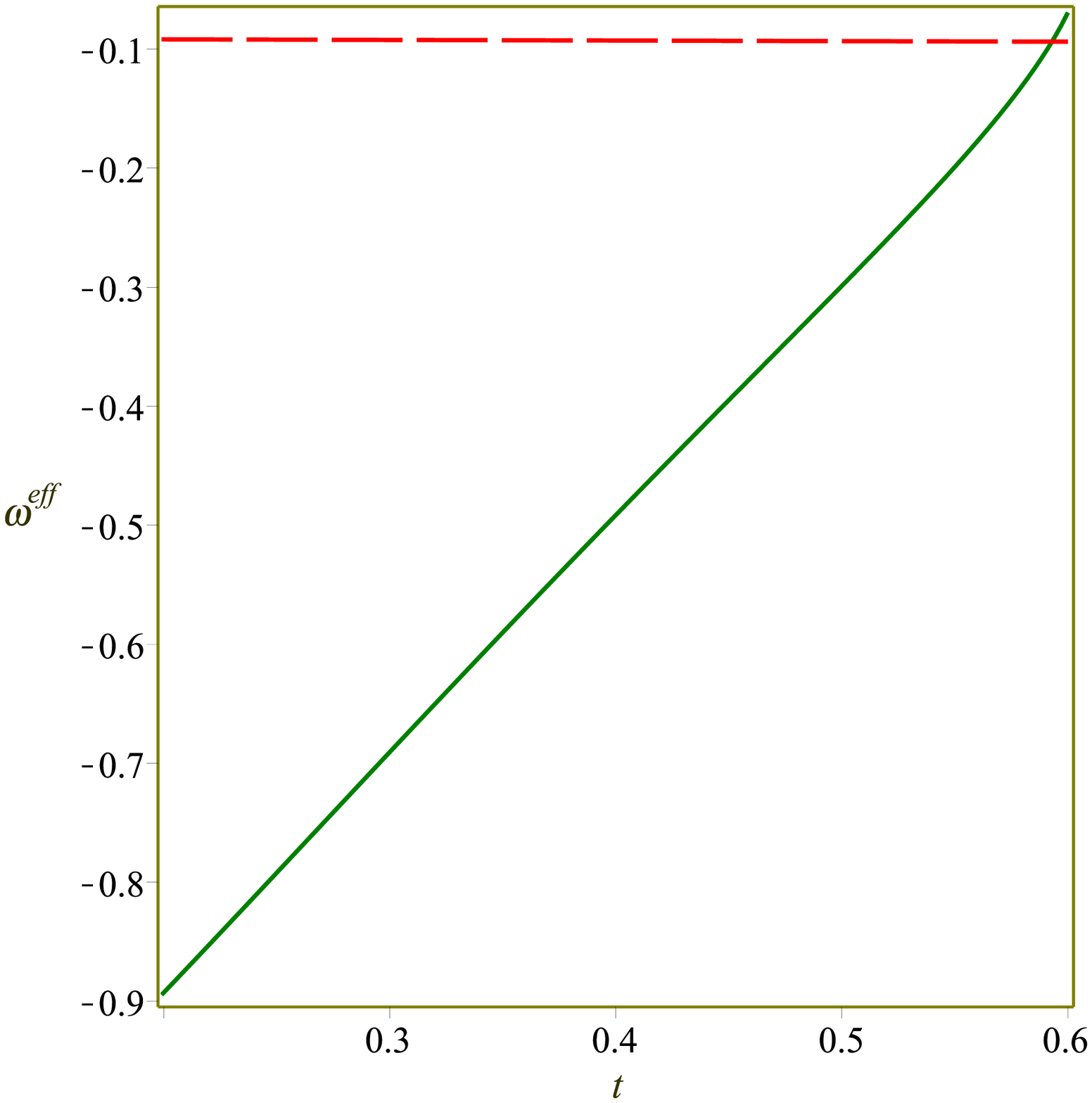,width=.5\linewidth} \caption{Behavior of $r-s$
(Left) and EoS (Right) parameters for $\epsilon=-1$ (line 1) and
$\epsilon=1$ (line 2).}
\end{figure}

We have considered the values of integration constants as
$c_{1}=-2$, $c_{3}=-10$, $c_{4}=10$ and $c_{5}=5.7$ to analyze the
graphical behavior of physical quantities.  Figure \textbf{1} shows
that the effective energy density is positively increasing for
$\epsilon=-1$ which manifests that our universe is in the expansion
phase. Figure \textbf{2} shows that the effective pressure and
deceleration parameter are negative for BT-III universe model which
support the current cosmic acceleration. Figure \textbf{3}
determines that $r-s$ and EoS parameters describe quintessence and
phantom phases of DE which represent the cosmic expansion. The
obtained solutions for $\epsilon=-1$ are consistent with recent
observations which indicate that this theory demonstrates expansion
of the universe.
\begin{figure}
\epsfig{file=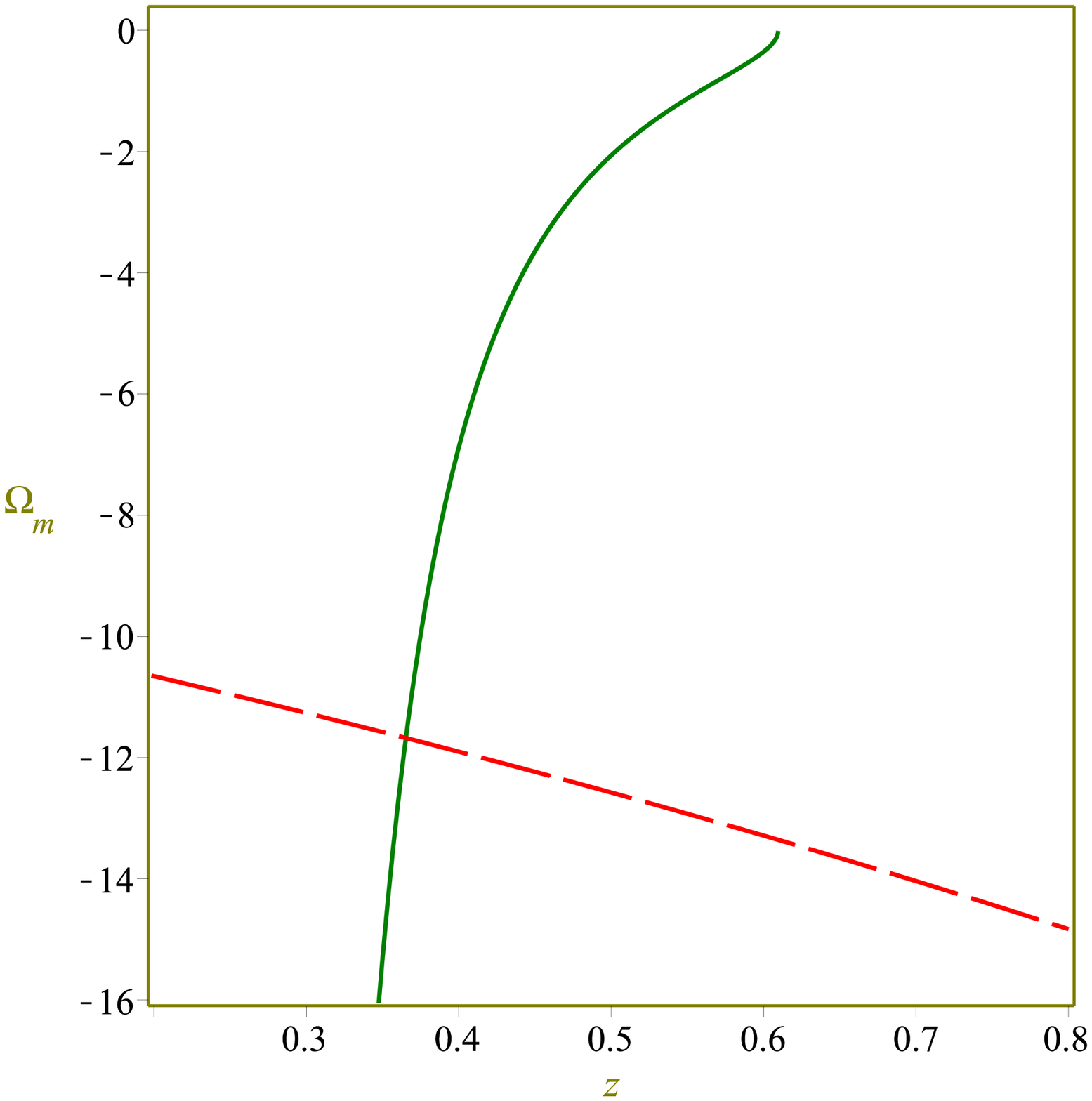,width=.5\linewidth}
\epsfig{file=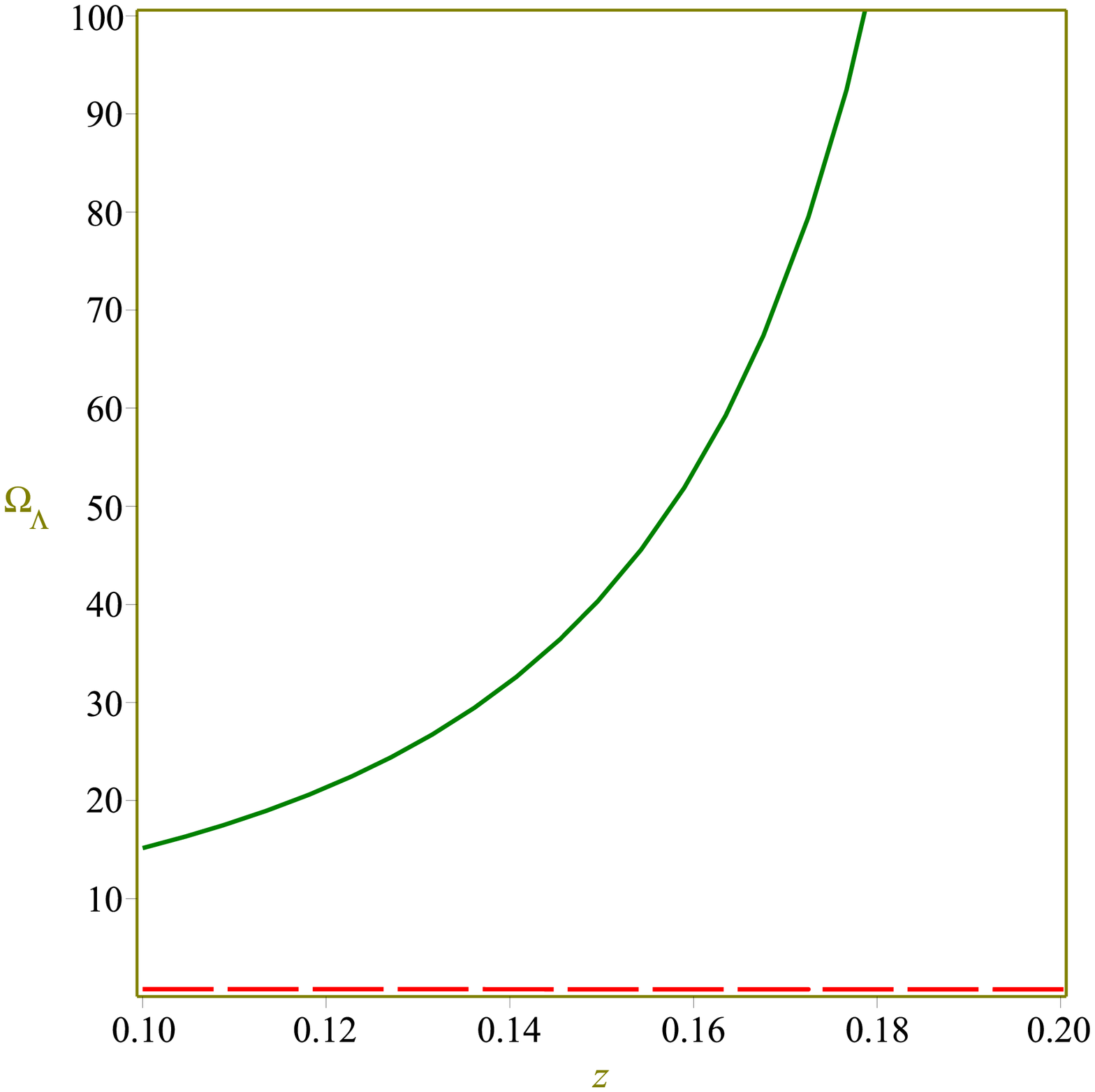,width=.5\linewidth} \caption{Plot of
$\Omega_{m}$ and $\Omega_{\Lambda}$ verses redshift parameter for
$\epsilon=-1$ (line 1) and $\epsilon=1$ (line 2).}
\end{figure}

The total amount of energy density can be expressed as fractional
energy density. The fractional density is defined as
\begin{equation}\nonumber
1+\Omega_{\sigma}=\Omega_{m}+\Omega_{\Lambda},
\end{equation}
where
\begin{eqnarray}\nonumber
\Omega_{m}=\frac{\rho}{3H^{2}}, \quad
\Omega_{\Lambda}=\frac{\rho_{\Lambda}}{3H^{2}}, \quad
\Omega_{\sigma}=\frac{\sigma^{2}}{3H^{2}}.
\end{eqnarray}
The evaluation of fractional densities corresponding to ordinary
matter $(\Omega_{m})$ and dark energy $(\Omega_{\Lambda})$ plays a
vital role to measure the contribution of these elements in the
cosmos. The densities for isotropic universe model defined as
$\Omega_{m}+\Omega_{\Lambda}=1$ whereas expression equality becomes
$\Omega_{m}+\Omega_{\Lambda}=1+\Omega_{\sigma}$ for anisotropic
universe model. We analyze the behavior of fractional densities
corresponding to matter and dark energy graphically at redshift
scale factor where $a(t)=a_{0}(1+z)^{-1}$ and $z$ is the redshift
parameter. From observations of Planck data 2018, it is suggested
that $\Omega_{m}\cong0.3111$ and $\Omega_{\Lambda}\cong0.6889$.
According to some recent observations, there are some evidences in
favor of closed universe model with fractional density
$\Omega_{\Lambda}\cong0.01$. For $\varepsilon=-1$, the fractional
density of matter indicates inconsistent behavior and the trajectory
of fractional density provides $\Omega_{m}=0.3$ for $\varepsilon=1$
as shown in Figure \textbf{4} (left plot). In this regard, it
implies consistent behavior with Planck data 2018. The right plot of
Figure \textbf{4} reveals the behavior of fractional density of dark
energy which shows consistent behavior with Plank data for
$\varepsilon=1$ and it exhibits inconsistent behavior for
$\varepsilon=-1$.

\section{Final Remarks}

Modified theories are assumed as the most propitious and elegant
proposals to examine the dark universe due to the presence of extra
higher-order geometric terms. In this paper, we have formulated
exact solutions of anisotropic and homogeneous spacetimes in
$f(R,T^{2})$ gravity. For this reason, we have considered the NS
technique to examine the exact solutions. We have formulated the
Lagrangian, NS generators with conserved values in the background of
EMSG. The behavior of exact solutions have been investigated through
different cosmological quantities. The main findings are summarized
as follows.
\begin{itemize}
\item We have established two non-zero NS generators and
corresponding conserved quantities. We have obtained the exact
solutions for BT-I, BT-III and KS universe models.
\item The effective energy density show accelerated and constant expansion
corresponding to BT-III, BT-I and KS spacetimes, respectively
(Figure \textbf{1}).
\item The value of effective pressure and deceleration parameter remain
negative for $\epsilon=-1$ which support the current cosmic
acceleration (Figure \textbf{2}).
\item The $r-s$ and
EoS parameters yield quintessence and phantom DE phases which
determine the rapid expansion of the universe (Figure \textbf{3}).
\item In the background of BT-III universe models, the analysis of
fractional density parameter of matter reveals that the EMSG is
consistent with Planck 2018 data. In case of KS universe model, this
consistency is not preserved (Figure \textbf{4}). We conclude that
the EMSG significantly explains the cosmic journey from decelerated
to accelerated epoch.
\end{itemize}
We find that first integrals of motion are very useful to obtain
viable cosmological solutions. It is found that the considered model
of EMSG supports the cosmic expansion. Extending this work to the
scalar field would be fascinating since it could provide a good
foundation for the investigation of the enigmatic cosmos in EMSG.

\vspace{0.25cm}

\section*{Appendix A}
\renewcommand{\theequation}{A\arabic{equation}}
\setcounter{equation}{0}

The values of $r-s$ parameters are
\begin{eqnarray}\nonumber
r&=&25\bigg[-126000c_{4}^{2}c_{5}^{2}c_{1}^{3}\epsilon^{3}c_{3}
(c_{5}+t)^{\frac{2}{3}}+94800c_{4}^{2}c_{1}^{3}\epsilon^{3}c_{3}
t^{2}(c_{5}+t)^{\frac{2}{3}}+7840c_{4}^{2}c_{1}^{2}\\\nonumber
&\times&c_{3}^{2}\epsilon^{2}t^{3}({c_{5}}+t)^{\frac{2}{3}}+928
c_{4}c_{5}^{5}c_{1}\epsilon
c_{3}^{3}\sqrt[3]{c_{5}+t}-11520{c_{4}}c_{5}^{4}c_{1}^{2}\epsilon
^{2}c_{3}^{2}\sqrt[3]{c_{5}+t}+c_{4}c_{5}^{3}\\\nonumber&\times&
34920c_{1}^{3}\epsilon^{3}c_{3}\sqrt[3]{c_{5}+t}-54000c_{4}c_{5}
c_{1}^{4}\epsilon^{4}t\sqrt[3]{c_{5}+t}-19080c_{4}c_{1}^{3}c_{3}
\epsilon^{3}t^{3}\sqrt[3]{c_{5}+t}\\\nonumber&-&960c_{4}c_{1}^{2}
c_{3}^{2}\epsilon^{2}t^{4}\sqrt[3]{c_{5}+t}+34560c_{4}c_{5}^{2}
c_{1}^{2}c_{3}^{2}\epsilon^{2}t^{2}\sqrt[3]{c_{5}+t}+1760c_{5}^{2}
c_{3}^{3}\epsilon t^{3}\sqrt[3]{c_{5}+t}\\\nonumber&\times&c_{4}
c_{1}-3240c_{4}c_{5}c_{1}^{3}c_{3}\epsilon^{3}t^{2}\sqrt[3]{c_{5}
+t}+21600c_{4}c_{5}c_{1}^{2}c_{3}^{2}\epsilon^{2}t^{3}\sqrt[3]
{c_{5}+t}+4480c_{4}c_{1}c_{3}^{3}\\\nonumber&\times&c_{5}\epsilon
t^{4}\sqrt[3]{c_{5}+t}-29280c_{4}^{2}c_{5}^{2}c_{1}^{2}c_{3}^{2}
\epsilon^{2}t(c_{5}+t)^{\frac{2}{3}}-31200c_{4}^{2}c_{5}c_{1}^{3}
c_{3}\epsilon^{3}t(c_{5}+t)^{\frac{2}{3}}-c_{1}^{2}\\\nonumber&
\times&39680c_{4}^{2}c_{5}c_{3}^{2}\epsilon^{2}t^{2}(c_{5}+t)^{
\frac{2}{3}}-3040c_{4}c_{5}^{4}c_{1}c_{3}^{3}\epsilon
t\sqrt[3]{c_{5}+t}+480c_{4}c_{5}^{3}c_{1}^{2}c_{3}^{2}\epsilon^{2}
t\sqrt[3]{c_{5}+t}\\\nonumber&-&5920c_{4}c_{5}^{3}c_{1}c_{3}^{3}
\epsilon
t^{2}\sqrt[3]{c_{5}+t}+50760c_{4}c_{5}^{2}c_{1}^{3}c_{3}\epsilon^{3}
t\sqrt[3]{c_{5}+t}+2025c_{5}^{3}c_{1}^{4}\epsilon^{4}-240c_{5}^{5}
c_{3}^{4}t^{2}\\\nonumber&+&160c_{5}^{4}c_{3}^{4}t^{3}+2025c_{1}^{4}
\epsilon^{4}t^{3}+480c_{5}^{3}c_{3}^{4}t^{4}+216c_{5}^{2}c_{3}^{4}
t^{5}-120c_{5}^{6}c_{3}^{4}t+16c_{3}^{4}c_{5}^{7}+c_{4}^{3}c_{5}^{2}
c_{1}^{3}\\\nonumber&\times&150400c_{3}\epsilon^{3}-153600c_{4}^{3}
c_{1}^{3}c_{3}\epsilon^{3}t^{2}-1620c_{5}c_{1}^{2}c_{3}^{2}\epsilon
^{2}t^{4}-648c_{5}c_{1}c_{3}^{3}\epsilon
t^{5}+165600c_{4}^{2}c_{5}\\\nonumber&\times&c_{1}^{4}\epsilon^{4}
(c_{5}+t)^{\frac{2}{3}}+165600c_{4}^{2}c_{1}^{4}\epsilon^{4}t(c_{5}
+t)^{\frac{2}{3}}-27000c_{4}c_{5}^{2}c_{1}^{4}\epsilon^{4}\sqrt[3]
{c_{5}+t}-27000c_{4}c_{1}^{4}\\\nonumber&\times&\epsilon^{4}t^{2}
\sqrt[3]{c_{5}+t}+360c_{5}^{5}c_{1}c_{3}^{3} \epsilon
t+2430c_{5}^{4}c_{1}^{2}c_{3}^{2}\epsilon^{2}t+2520c_{5}^{4}c_{1}
c_{3}^{3}\epsilon
t^{2}-8100c_{5}^{3}c_{1}^{3}c_{3}\epsilon^{3}t\\\nonumber&-&3240
c_{5}^{3}c_{1}^{2}c_{3}^{2}\epsilon^{2}t^{2}+1920c_{5}^{3}c_{1}
c_{3}^{3}\epsilon
t^{3}-4050c_{5}^{2}c_{1}^{3}c_{3}\epsilon^{3}t^{2}-540c_{5}^{2}
c_{1}c_{3}^{3}\epsilon
t^{4}+2700c_{5}c_{1}^{3}\\\nonumber&\times&c_{3}\epsilon^{3}t^{3}
-5940c_{5}^{2}c_{1}^{2}c_{3}^{2}\epsilon^{2}t^{3}-456000c_{4}^{3}
c_{5}c_{1}^{4}\epsilon^{4}-456000c_{4}^{3}c_{1}^{4}\epsilon^{4}t
+6075c_{5}^{2}c_{1}^{4}\epsilon^{4}t\\\nonumber&-&3375c_{3}
c_{5}^{4}c_{1}^{3}\epsilon^{3}+6075c_{5}c_{1}^{4}\epsilon^{4}t^{2}
+2025c_{1}^{3}c_{3}\epsilon^{3}t^{4}+486c_{1}^{2}c_{3}^{2}\epsilon
^{2}t^{5}-348c_{5}^{6}c_{1}c_{3}^{3}\epsilon+c_{4}^{4}c_{1}^{4}
\\\nonumber&\times&448000\epsilon^{4}\sqrt[3]{c_{5}+t}+1836c_{5}
^{5}c_{1}^{2}c_{3}^{2}\epsilon^{2}+768c_{4}c_{1}c_{3}^{3}\epsilon
t^{5}\sqrt[3]{c_{5}+t}+18240c_{4}^{2}c_{5}^{3}c_{1}^{2}c_{3}^{2}
\epsilon^{2}\\\nonumber&\times&(c_{5}+t)^{\frac{2}{3}}-3200a{1}^{3}
c_{5}c_{1}^{3}c_{3}\epsilon^{3}t\bigg]\bigg[(4c_{5}c_{3}(c_{5}+t)
^{\frac{2}{3}}-\sqrt[3]{c_{5}+t}(15c_{1}\epsilon(c_{5}+t)^{\frac{2}
{3}}\\\nonumber&-&6c_{3}t(c_{5}+t)^{\frac{2}{3}}+40c_{4}c_{1}\epsilon)
^{4}\bigg]^{-1},\\\nonumber
s&=&\frac{1}{6}\bigg[432000c_{4}^{3}c_{5}c_{1}^{3}c_{3}\epsilon^{3}
t-1998000c_{4}^{2}c_{5}^{2}c_{1}^{3}c_{4}\epsilon^{3}(c_{5}+t)^{\frac{2}
{3}}+1053000c_{4}c_{5}^{2}c_{1}^{3}c_{3}\\\nonumber&\times&\epsilon^{3}
t\sqrt[3]{c_{5}+t}-103200c_{4}c_{5}^{3}c_{1}^{2}c_{3}^{2}\epsilon^{2}t
\sqrt[3]{c_{5}+t}-135200c_{4}c_{5}^{3}c_{1}c_{3}^{3}\epsilon
t^{2}\sqrt[3]{c_{5}+t}\\\nonumber&-&876800c_{4}^{2}c_{5}c_{1}^{2}c_{3}
^{2}\epsilon^{2}t^{2}(c_{5}+t)^{\frac{2}{3}}-1356000c_{4}^{2}c_{5}c_{1}
^{3}c_{3}\epsilon^{3}t(c_{5}+t)^{\frac{2}{3}}-50400c_{4}c_{1}\\\nonumber&
\times&c_{5}^{4}c_{3}^{3}\epsilon
t\sqrt[3]{c_{5}+t}-13600c_{4}c_{5}^{2}c_{1}c_{3}^{3}\epsilon
t^{3}\sqrt[3]{c_{5}+t}+783000c_{4}c_{5}c_{1}^{3}c_{3}\epsilon^{3}t^{2}
\sqrt[3]{c_{5}+t}\\\nonumber&+&712800c_{4}c_{5}c_{1}^{2}c_{3}^{2}
\epsilon^{2}t^{3}\sqrt[3]{c_{5}+t}+112000c_{4}c_{5}c_{1}c_{3}^{3}\epsilon
t^{4}\sqrt[3]{c_{5}+t}-2232c_{5}^{6}c4^{4}t\\\nonumber&-&424800c_{4}^{2}
c_{5}^{2}c_{1}^{2}c_{3}^{2}\epsilon^{2}t(c_{5}+t)^{\frac{2}{3}}+547200c_{4}
c_{5}^{2}c_{1}^{2}c_{3}^{2}\epsilon^{2}t^{2}\sqrt[3]{c_{5}+t}+1440c_{5}^{4}
c_{3}^{4}t^{3}\\\nonumber&-&5616c_{5}^{5}c_{3}^{4}t^{2}+12240c_{5}^{3}c_{3}
^{4}t^{4}+8424c_{5}^{2}c_{3}^{4}t^{5}-432c_{5}c_{3}^{4}t^{6}-1296c_{3}^{4}
t^{7}+144c_{3}^{4}c_{5}^{7}\\\nonumber&-&121500c_{5}c_{1}^{3}c4\epsilon
^{3}t^{3}-121500c_{5}^{2}c_{1}^{2}c_{3}^{2}\epsilon^{2}t^{3}+8100c_{5}^{2}
c_{1}c_{3}^{3}\epsilon
t^{4}-182250c_{5}^{2}c_{1}^{3}c_{3}\epsilon^{3}t^{2}\\\nonumber&+&64800c_{5}
^{3}c_{1}c_{3}^{3}\epsilon
t^{3}-121500c_{5}^{3}c_{1}^{3}c_{3}\epsilon^{3}t+60750c_{5}^{4}c_{1}^{2}
c_{3}^{2}\epsilon^{2}t+48600c_{5}^{4}c_{1}c_{3}^{3}\epsilon
t^{2}-c_{1}^{4}\\\nonumber&\times&135000c_{4}\epsilon^{4}t^{2}\sqrt[3]{c_{5}
+t}+3240c_{5}^{5}c_{1}c_{3}^{3}\epsilon
t-135000c_{4}c_{5}^{2}c_{1}^{4}\epsilon^{4}\sqrt[3]{c_{5}+t}+1980000\\\nonumber
&\times&c_{4}^{2}c_{5}c_{1}^{4}\epsilon^{4}(c_{5}+t)^{\frac{2}{3}}+1980000
c_{4}^{2}c_{1}^{4}\epsilon^{4}t(c_{5}+t)^{\frac{2}{3}}-121500c_{5}c_{1}^{2}
c_{3}^{2}\epsilon^{2}t^{4}-29160\\\nonumber&\times&c_{5}c_{1}c_{3}^{3}\epsilon
t^{5}+2736000c_{4}^{3}c_{5}^{2}c_{1}^{3}c_{3}\epsilon^{3}-2304000c_{4}^{3}c_{1}
^{3}c_{3}\epsilon^{3}t^{2}+24300c_{5}^{5}c_{1}^{2}c_{3}^{2}\epsilon^{2}-c_{1}
^{2}c_{3}^{2}\\\nonumber&\times&36450\epsilon^{2}t^{5}+8640000c_{4}^{4}c_{1}
^{4}\sqrt[3]{c_{5}+t}\epsilon^{4}-4860c_{5}^{6}c_{1}c_{3}^{3}\epsilon-30375
c_{1}^{3}c_{3}\epsilon^{3}t^{4}-c_{5}^{4}c_{1}^{3}c_{3}\\\nonumber&\times&
30375\epsilon^{3}-7560000c_{4}^{3}c_{5}c_{1}^{4}\epsilon^{4}-7560000c_{4}^{3}
c_{1}^{4}\epsilon^{4}t-12960c_{1}c_{3}^{3}\epsilon
t^{6}+642000c_{4}^{2}c_{1}^{3}\\\nonumber&\times&c_{3}\epsilon^{3}t^{2}
(c_{5}+t)^{\frac{2}{3}}-149600c_{4}^{2}c_{1}^{2}c_{3}^{2}\epsilon^{2}t^{3}
(c_{5}+t)^{\frac{2}{3}}+12960c_{4}c_{5}^{5}c_{1}c_{3}^{3}\epsilon\sqrt[3]
{c_{5}+t}-c_{4}\\\nonumber&\times&172800c_{5}^{4}c_{1}^{2}c_{3}^{2}\epsilon^{2}
\sqrt[3]{c_{5}+t}+441000c_{4}c_{5}^{3}c_{1}^{3}c_{3}\epsilon^{3}\sqrt[3]{c_{5}
+t}-270000c_{4}\epsilon^{4}t\sqrt[3]{c_{5}+t}\\\nonumber&\times&c_{5}c_{1}^{4}
+235200c_{4}c_{1}^{2}c_{3}^{2}\epsilon^{2}t^{4}\sqrt[3]{c_{5}+t}+171000c_{4}
c_{1}^{3}c_{3}\epsilon^{3}t^{3}\sqrt[3]{c_{5}+t}+53760c_{4}c_{1}c_{3}^{3}
\\\nonumber&\times&\epsilon t^{5}\sqrt [3]{c_{5}+t}+302400c_{4}^{2}c_{5}^{3}
c_{1}^{2}c_{3}^{2}\epsilon^{2}(c_{5}+t)^{\frac{2}{3}}\bigg]\bigg[\bigg\{600c_{4}
^{2}c_{1}^{2}\epsilon^{2}\sqrt[3]{c_{5}+t}+21(c_{5}+t)^{\frac{2}{3}}\\\nonumber&
\times&c_{5}^{3}c_{3}^{2}-45c_{5}^{2}c_{1}c_{3}\epsilon(c_{5}+t)^{\frac{2}{3}}
+33c_{5}^{2}c_{3}^{2}t(c_{5}+t)^{\frac{2}{3}}+3c_{5}c_{3}^{2}t^{2}(c_{5}+t)
^{\frac{2}{3}}-(c_{5}+t)^{\frac{2}{3}}\\\nonumber&\times&90c_{5}c_{1}c_{3}\epsilon
t-45c_{1}c_{3}(c_{5} +t)^{\frac{2}{3}}\epsilon
t^{2}-9c_{3}^{2}(c_{5}+t)^{\frac{2}{3}}t^{3}+270c_{4}c_{5}^{2}c_{1}c_{3}\epsilon
-75c_{4}c_{5}c_{1}^{2}
\epsilon^{2}\\\nonumber&+&490c_{4}c_{5}c_{1}c_{3} \epsilon
t-75c_{4}c_{1}^{2}\epsilon^{2}t+220 c_{4}c_{1}c_{3}\epsilon
t^{2}\bigg\}\bigg\{4c_{5}c_{3}(c_{5}+t)^{\frac{2}{3}}-15(c_{5}+t)^{\frac{2}{3}}
c_{1}\epsilon\\\nonumber&-&6c_{3}t(c_{5}+t)^{\frac{2}{3}}+40c_{4}c_{1}\epsilon
\bigg\}^{2}\bigg]^{-1}.
\end{eqnarray}

\end{document}